\documentclass[aps,prb,preprint,superscriptaddress]{revtex4-2}

\usepackage{dcolumn}   
\usepackage{bm}        
\usepackage{amsmath}
\usepackage{amssymb}
\usepackage[utf8]{inputenc}
\usepackage{graphicx,siunitx}
\usepackage{amsmath}
\DeclareUnicodeCharacter{2212}{-}
\usepackage[english]{babel}
\usepackage{booktabs}
\usepackage[svgnames]{xcolor}
\usepackage[export]{adjustbox}
\usepackage[percent]{overpic}
\usepackage[export]{adjustbox}
\usepackage{array}
\usepackage[style=base]{caption}
\DeclareSIUnit\angstrom{\text {Å}}
\begin{document}
	\title{Pressure-induced softening in bulk modulus due to magneto-elastic coupling in Nd$_2$CoFeO$_6$ double Perovskite}
	
	\author{Bidisha Mukherjee}
	\affiliation{National Centre for High-Pressure Studies, Department of Physical Sciences, Indian Institute of Science Education and Research Kolkata, Mohanpur Campus, Mohanpur 741246, Nadia, West Bengal, India.}
    \author{Mrinmay Sahu}
	\affiliation{National Centre for High-Pressure Studies, Department of Physical Sciences, Indian Institute of Science Education and Research Kolkata, Mohanpur Campus, Mohanpur 741246, Nadia, West Bengal, India.}
    \author{Debabrata Samanta}
	\affiliation{National Centre for High Pressure Studies, Department of Physical Sciences, Indian Institute of Science Education and Research Kolkata, Mohanpur Campus, Mohanpur 741246, Nadia, West Bengal, India.}
    \author{Bishnupada Ghosh}
	\affiliation{National Centre for High Pressure Studies, Department of Physical Sciences, Indian Institute of Science Education and Research Kolkata, Mohanpur Campus, Mohanpur 741246, Nadia, West Bengal, India.}
    \author{Boby Joseph}
	\affiliation{Elettra-Sincrotrone Trieste, S. S. 14 km 163.5, 34149 Basovizza, Trieste, Italy}
	\author{Goutam Dev Mukherjee}
	\email [Corresponding author: ]{goutamdev@iiserkol.ac.in}
	\affiliation{National Centre for High-Pressure Studies, Department of Physical Sciences, Indian Institute of Science Education and Research Kolkata, Mohanpur Campus, Mohanpur 741246, Nadia, West Bengal, India.}
	\date{\today}
	
	\begin{abstract}
  Double perovskite oxide materials have garnered tremendous interest due to their strong spin-lattice-charge coupling. Interesting in their own right, rare-earth-based DPOs have yet to be subjected to high-pressure studies. In this paper,
   we have investigated the structural response of Nd$_2$CoFeO$_6$ to pressure by XRD and Raman spectroscopic measurements. From XRD data, we have observed pressure-induced structural transition from the orthorhombic phase to the monoclinic phase at about 13.8~\si{\giga\pascal}. An anomalous increase in compressibility at a much lower pressure($\sim$1.1~\si{\giga\pascal}) is seen where no structural transition occurs. At about the same pressure, a sudden drop in the slope of Raman modes is observed. Further investigation at low temperatures reveals that the B$_g$ Raman mode is strongly affected by magnetic interactions. Additional high-pressure Raman experiments with the application of a magnetic field indicated that the mentioned anomaly around 1.1~\si{\giga\pascal} can be explained by a high-spin to low-spin transition of Co$^{3+}$.   
   
	\end{abstract} 
	\maketitle
	\section{INTRODUCTION}
Double perovskite oxide (DPO) class of materials with the chemical form A$_2$BB$'$O$_6$ (where, A = alkaline earth or rare-earth metal, B and B$'$ = transition metal) have drawn the attention of the global scientific community because of their potential technological relevance and interesting physical properties \cite{vasala2015a2b}. In particular, the Re$_2$BB$'$O$_6$ family, where Re is rare earth element, is reported to possess multiple magnetic and electrical properties because of their exceptional structural and compositional flexibility \cite{vasala2015a2b}. These low-cost materials are easy to fabricate and show high thermoelectric figure-of-merit \cite{tanwar2019enhancement}, spin reorientation transitions \cite{shukla2019low,haripriya2017temperature,lohr2018multiferroic}, high chemical stability, magnetocaloric effect \cite{murthy2015giant, das2019nonlinear}, spin canting \cite{shuklamagnetic}, multiferroicity \cite{lohr2018multiferroic}, magnetic entropy change, \cite{das2019nonlinear, dong2020structural} Griffith’s-like phase \cite{pal2019b} etc. However, the effect of lattice strain on these DPOs is yet unclear because of a lack of systematic studies in this direction. Among this class of materials, Nd$_2$CoFeO$_6$ (NCFO) shows insulating behaviour at room temperature. It has a paramagnetic state above 541~K and transforms to an antiferromagnetic state below 246~K. Strong short-range interactions dominate at intermediate temperatures \cite{de2020insulator}. 

The effect of pressure at ambient temperature has been studied for a few double perovskite oxides. Manoun et al. \cite{manoun2008high} have reported two structural phase transitions in Sr$_2$CoWO$_6$: from tetragonal (I4/m) to monoclinic ($P2_1/n$) at 2.2~\si{\giga\pascal} and monoclinic to amorphous at 12.7~\si{\giga\pascal}. High-pressure Raman analysis suggests a possible phase transition at 3.2~\si{\giga\pascal} and 2 new modes appear above 10.4~\si{\giga\pascal} in Sr$_2$CaWO$_6$ \cite{manoun2004high}. In Sr$_2$FeMoO$_6$ a spin cross-over of Fe ion from high spin to low spin state is found at 33-36~\si{\giga\pascal} accompanied by ferrimagnetic half-metallic to non-magnetic semiconductor state and above 45~\si{\giga\pascal} the non-magnetic compound behaves as metallic \cite{qian2012effect}.
In Re$_2$CoMnO$_6$, magnetic interactions are largely dependent on the
Mn–O–Mn bond angles and Co--O bond length as suggested by Troyanchuk et al. \cite{troyanchuk2000magnetic}. Changes in the octahedral distortion in most double-perovskite oxides give us the ability to control important dielectric, electric, magnetic, piezoelectric, and magnetoresistive properties of these materials. In particular, the distortion may be enhanced or diminished depending on the relative compressibilities of the AO$_{12}$ and BO$_6$ polyhedra. These distortions in octahedra can be achieved by changing pressure, temperature, and ionic radius of cation \cite{vasala2015a2b}. It can be expected that NCFO may also significantly change its polyhedral properties, bond length, bond angle, and hence structural, electronic, and magnetic properties upon applying pressure. de Oliveira et al. \cite{de2020insulator} have shown insulator to metal transition of NCFO with increasing temperature. From temperature-dependent X-Ray Diffraction (XRD) measurements, they observed a step in the $a$-lattice parameter at about 825~K and reported it as evidence of spin-lattice coupling. Several ab-initio first principle calculations have predicted novel phase transition on the effect of applying pressure in this class of materials. Zhao et al. \cite{zhao2015predicted} have predicted a high spin to low spin state transition of Co$^{2+}$ accompanied by a significant volume collapse as well as a change in electronic properties in Re$_2$CoMoO$_6$. La$_2$VMnO$_6$ becomes half-metallic ferrimagnet from insulating ferrimagnet upon compressing volume \cite{zu2013pressure}. Still, the structural evolution of any rare-earth-based DPO with pressure has not been investigated by XRD or Raman spectroscopic measurements to the best of our knowledge.
 
In this study, the conventional solid-state reaction approach was used to synthesize polycrystalline NCFO sample. Pressure dependent XRD study was performed up to 18.8~\si{\giga \pascal} to investigate any structural changes in the sample.  Pressure dependent Raman spectroscopy measurements were carried out up to 27.4~\si{\giga \pascal}. We report structural transition from orthorhombic $Pnma$ to monoclinic $P2_1/n$ structure at $\sim14.8$~\si{\giga\pascal}, with anomalies observed around 1 to 2~\si{\giga \pascal} range in structural parameters. Temperature-dependent Raman spectroscopy measurement (from 22~K to room temperature) reveals that anomalies in Raman mode frequency are associated with changes in the magnetic properties of the sample. A comparison of low-temperature Raman spectroscopic data and high-pressure Raman data with and without the application of a magnetic field reveals a strong magneto-elastic coupling in the sample.
	
	\section{EXPERIMENTAL SECTION}
NCFO powder was synthesized with high-purity raw materials using the solid-state synthesis method prescribed by de Oliveira et al. \cite{de2020insulator} and characterized for phase purity using Lab-XRD at a wavelength of 1.0546~Å.
High-pressure XRD and Raman spectroscopic measurements were performed using a piston-cylinder type diamond anvil cell (DAC) with a culet diameter of \SI{300}{\micro\metre}. A \SI{290}{\micro\metre} thick steel gasket was indented to a thickness of \SI{45}{\micro\metre}. At the centre of the indented region, a hole of diameter \SI{100}{\micro\metre} was drilled through the gasket using an electric discharge machine. The gasket was then placed on the lower diamond and a very small amount of sample along with pressure transmitting medium (PTM) and pressure calibrant were loaded into the central hole. $4:1$ methanol-ethanol mixture was used as the PTM and the Ruby fluorescence technique was used for pressure calibration \cite{mao1986calibration}. A small amount of fine silver powder was also added along with the sample for pressure calibration in high-pressure XRD measurements.

The Raman spectra of the sample were taken with a Monovista confocal micro-Raman system from S\&I GmbH equipped with a Cobolt Samba \SI{532}{\nano\metre} diode-pumped laser. A long working distance infinitely corrected $20\times$ objective was used to focus the laser beam and also collect the scattered signal from the sample using back-scattering geometry. The collected scattered light was dispersed using a grating with 1500 grooves/mm having a spectral resolution of about \SI{1.2}{\centi\metre}$^{-1}$. An edge filter for Rayleigh line rejection was used.

High-pressure XRD measurements were carried out at the XPRESS beamline in the ELETTRA synchrotron source in Trieste, Italy.  A monochromatic x-ray beam with a wavelength of 0.4957 Å was incident on the sample with a collimated beam diameter of about \SI{50}{\micro\metre}. The diffraction patterns were recorded using a
PILATUS 3S 6M detector. The distance from the sample to the detector was calibrated using the XRD pattern of a standard sample CeO$_2$. Diffraction patterns were integrated to $2\theta$ vs intensity profile using DIOPTAS software \cite{prescher2015dioptas}. To analyse the XRD data, GSAS \cite{toby2001expgui}, PCW \cite{kraus1996powder}, EoSfit7 \cite{gonzalez2016eosfit7} and Vesta \cite{momma2008vesta} software were used.


Raman spectra measurements at low-temperature were taken using a closed cycle cryostat. A Cernox sensor, placed in the vicinity of the sample, measured its temperature, while a Lakeshore 325 controller was used to set and control the temperature. The scattered light is collected using a set of appropriate optics which couple the micro-Raman spectrometer to the cryostat. The sample was mounted on a copper sample holder, attached to the cold head.


For high-pressure Raman measurements under a magnetic field, we have used a miniature Copper-Beryllium (opposing plate type) DAC having a culet size of 600~\si{\micro\metre}. This cell can be used for a maximum pressure of about 8~\si{\giga\pascal}. For the generation of a magnetic field, we have used a double pole electromagnet for a maximum magnetic field of 0.46T at the center of the poles with a separation of 1.5~\si{\centi\metre} below the microscope objective of the Raman setup. All these Raman experiments were carried out by LabRam HR 800 Micro Raman facility with excitation laser of wavelength 488~\si{\nano\metre} and using a grating with 1800~grooves/mm.


	\section{RESULTS AND DISCUSSION}
\subsection{Sample Synthesis and Characterization}
\begin{figure}[th!]
    \centering
    \begin{overpic}[trim=37 30 40 10,clip,width=0.6\linewidth]{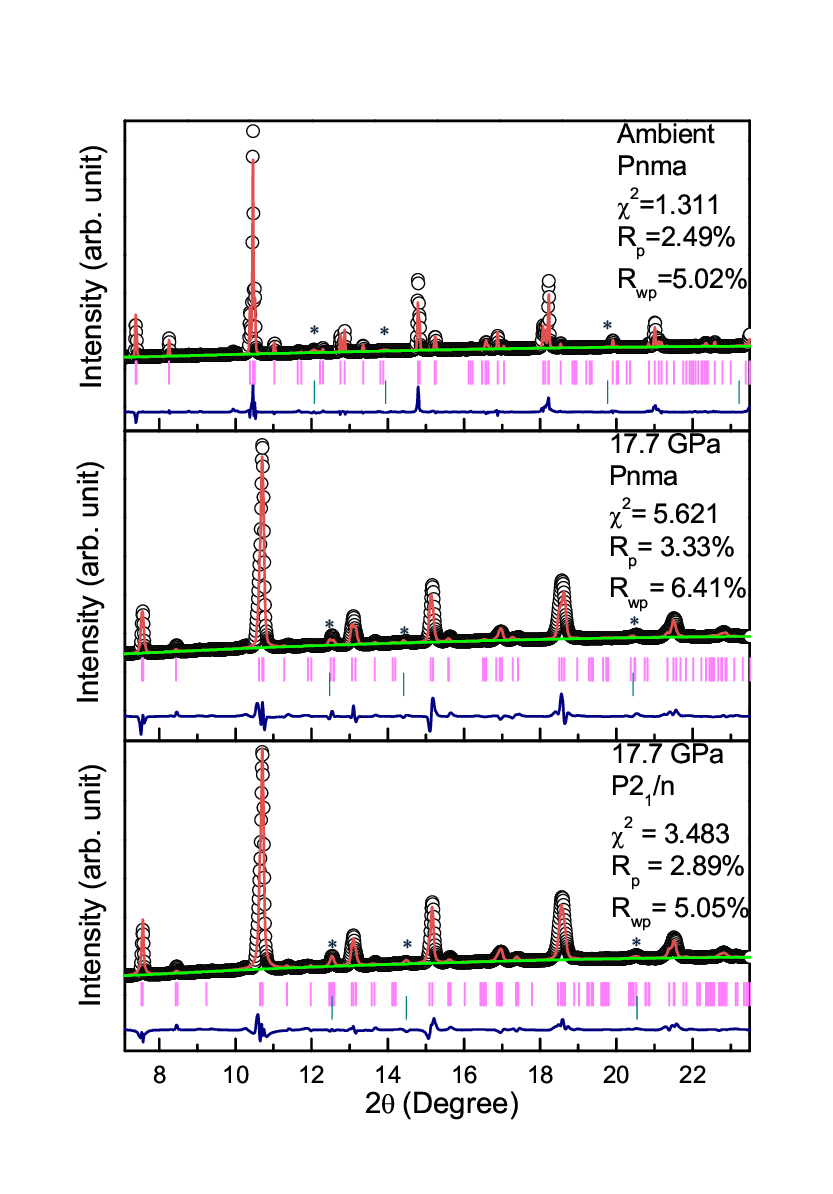} 
    \put (6,88) {(a)}
    \put (6,59) {(b)}
    \put (6,32) {(c)}
    \end{overpic}  \hfill  \begin{minipage}{0.37\linewidth}
    \vspace{-15cm}
    \begin{overpic}[width=\linewidth]{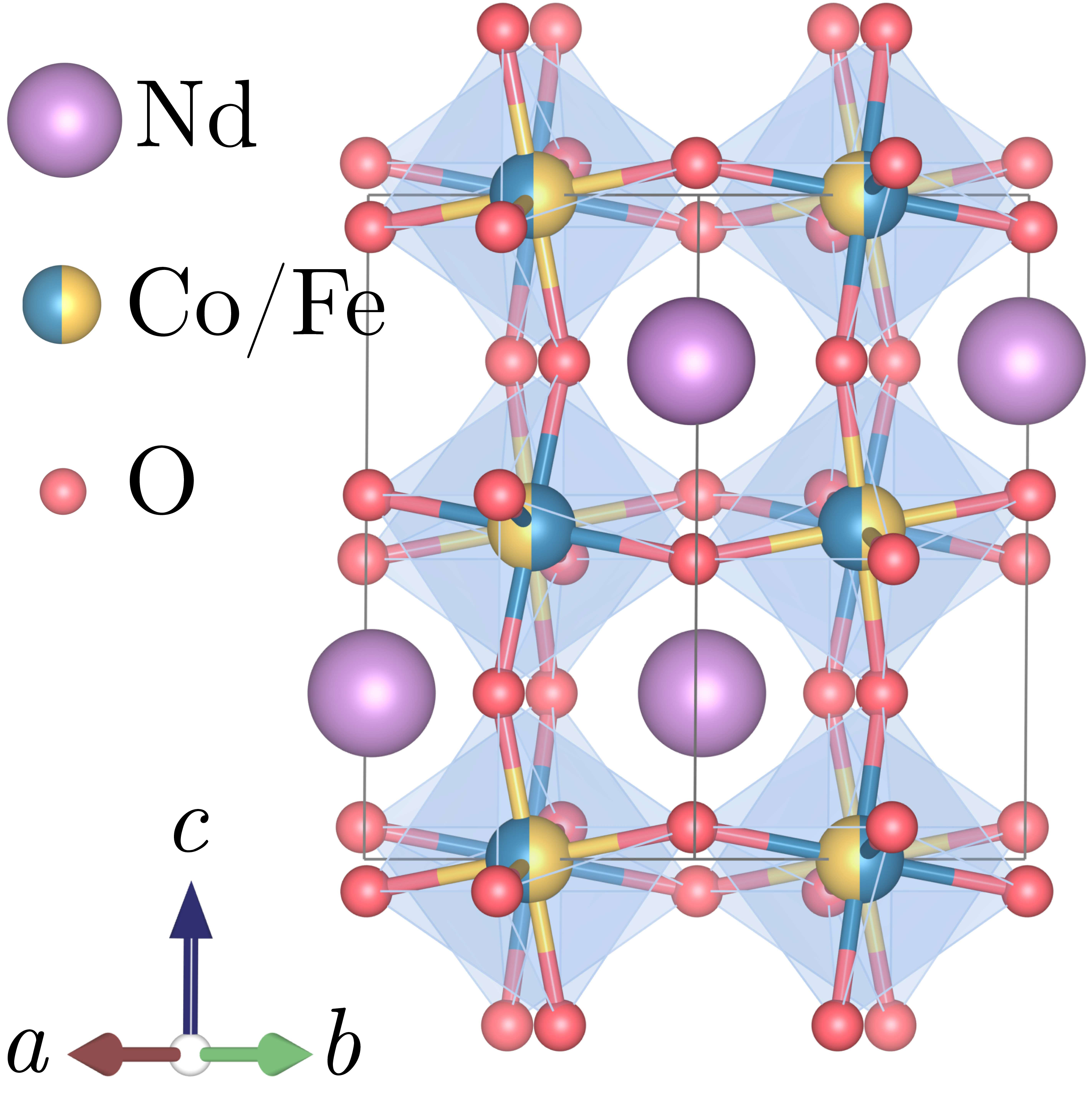} 
    \put (1,96) {(d)}
    \end{overpic} \\
    \vspace{2cm}
    \begin{overpic}[width=\linewidth]{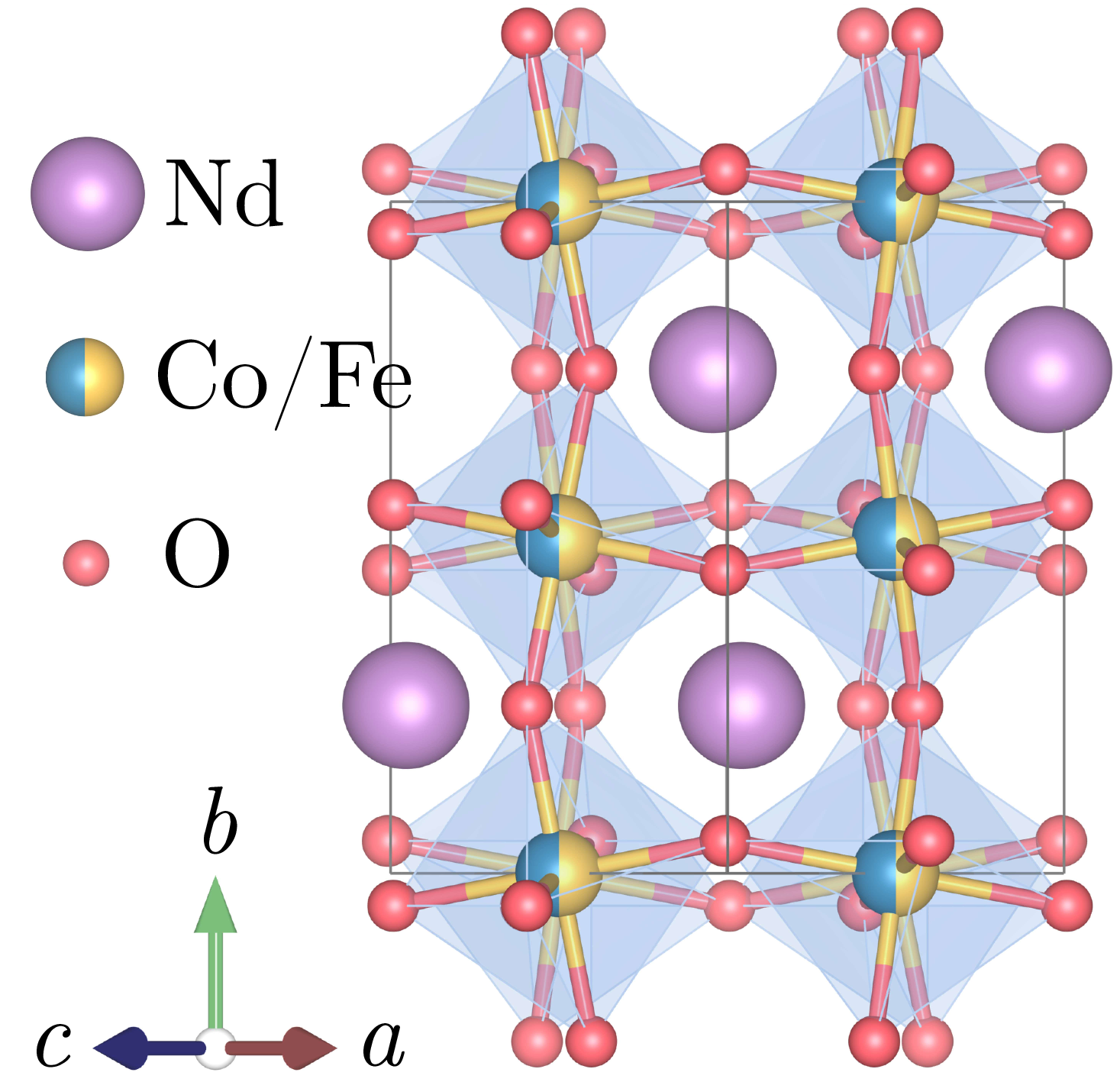} 
        \put (1,100) {(e)}
        \end{overpic}
    \end{minipage}
     
        \caption{ Rietveld refinement of the XRD pattern at (a) ambient, (b) 15.6~\si{\giga\pascal} and (c) 17.7~\si{\giga\pascal}. Black circles represent experimental data. Red, green, and navy lines are Rietveld fit to the experimental data, background, and difference between experimental and calculated data, respectively. The magenta and dark-cyan vertical lines show the Bragg peaks of the sample and silver, respectively. The silver peaks are marked with asterisks (*). Schematic representations of the unit cell at (d) ambient conditions and (e) 17.7~\si{\giga\pascal} are shown on the right side.}
        \label{fig:ambrefinementMono17.7}
 \end{figure}

The synthesized powder sample is characterized by XRD measurement. The ambient XRD pattern (Fig.~\ref{fig:ambrefinementMono17.7}(a)) matches the orthorhombic structure reported by de Oliveira et al. \cite{de2020insulator}. The refined lattice parameters obtained from our sample are $a=5.47899(9)$~\si{\angstrom}, $b=7.6804(1)$~\si{\angstrom}, $c=5.4116(1)$~\si{\angstrom} and volume = 227.727(7)~\si{\angstrom}$^3$. The orthorhombic structure has $Pnma$ space group (Space group no. 62) and all the structural details agree with the reported literature~\cite{de2020insulator}. The Rietveld refinement of the XRD pattern is carried out by using the atomic positions reported by Dhilip et al.~\cite{dhilip2022novel} for Sm$_2$CoFeO$_6$ material. As oxygen atom positions are insensitive to the peak intensities obtained from powder XRD measurements, we have not refined the oxygen atom positions. Fig.~\ref{fig:ambrefinementMono17.7}(a) shows the Rietveld refinement fit of our XRD pattern at ambient conditions which is excellent considering the good $R_p$ and $R_{wp}$ values. The stability of a double perovskite structure can be determined using the Goldschmidt tolerance factor, $t_f$, given by
\begin{equation}
    t_f=\frac{r_A  + r_O }{\sqrt{2}\left ( 
     \frac{r_B+r'_B}{2} +  r_O  \right )}
\end{equation}
where $r_A$, $r_B$, $r_{B}'$, $r_O$ are Shannon radii \cite{shannon1976revised} of A, B, B$'$ and O ions respectively. The calculated value of $t_f$ for NCFO is less than 0.97, which is also in good agreement with orthorhombic structure \cite{lufaso2006structure}. A schematic representation of a unit cell of NCFO is shown in Fig.~\ref{fig:ambrefinementMono17.7}(d) and the relative atomic positions after Rietveld refinement are given in Table~\ref{table:1}.
\begin{table}[ht!] 
\begin{center}
\begin{tabular}{ |c|c|c|c|c|c| } 
 \hline
 atom & Wyckoff position & x/a & y/b & z/c\\
 \hline
 Nd & 4c & 0.0437(4) & 0.25 & 0.9984(28)\\ 
 Fe/Co & 4b & 0.0 & 0.0 & 0.5\\
 O & 4c & 0.4842 & 0.25 & 0.07986\\
 O & 8d & 0.70552 & 0.95177 & 0.30209\\
 \hline
\end{tabular}\\
\end{center}
 \caption{Relative atomic positions at ambient condition after Rietveld refinement}
\label{table:1}
\end{table}

\subsection{High Pressure XRD}
\begin{figure}[ht!]
     \centering
     
     \includegraphics[width=0.75\linewidth]{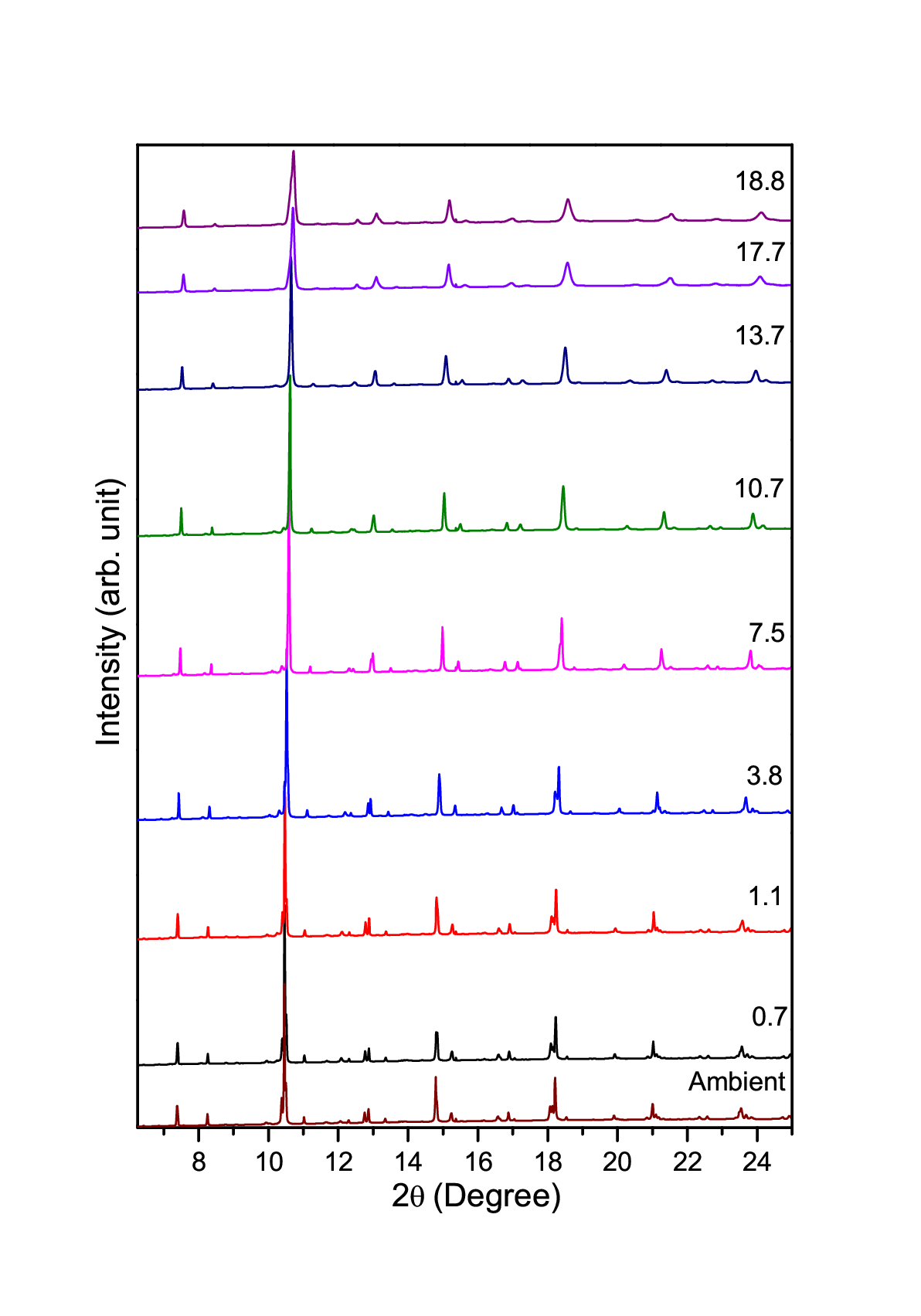}
     \caption{Pressure evolution of XRD pattern of NCFO up to 18.8~\si{\giga\pascal}. Pressures (in \si{\giga\pascal}) are indicated above the respective XRD plots.}
     \label{fig:pressevoXRD}
 \end{figure}

The pressure evolution of the XRD pattern at selected pressures is shown in Fig.~\ref{fig:pressevoXRD}. Rietveld refinement of the high-pressure XRD data reveals that orthorhombic structure gives a good fit up to about ~13.6~\si{\giga \pascal}. At and above 14.8~\si{\giga \pascal} the XRD pattern could not be fitted well to the orthorhombic structure. Indexing the XRD pattern at 17.7 GPa, the highest pressure of our XRD measurements, gives a monoclinic cell with high figure of merit with lattice parameters, $a=5.3092(9)$~\si{\angstrom}, $b=5.3139(9)$~\si{\angstrom}, $c=7.5512(7)$~\si{\angstrom}, $\beta=90.503(7)$ and volume = 213.03~\si{\angstrom}$^3$.  Considering group-subgroup changes this can be matched to space group $P2_1/n$ (Space group no. 14). The atomic positions in the unit cell of the monoclinic structure are estimated from the parent structure using the PCW software \cite{kraus1996powder}. Rietveld refinements using these atomic positions give a very good fit to the obtained XRD pattern (Fig.~\ref{fig:ambrefinementMono17.7}(c)). The refined atomic positions at 17.7~\si{\giga \pascal} are given in Table~\ref{table:2}. In Fig.~\ref{fig:ambrefinementMono17.7}(b) we have shown the Rietveld refinement of the XRD pattern at 17.7~\si{\giga\pascal} using the parent orthorhombic phase, which is found to be poor in comparison to the monoclinic phase considering the higher values of $R_p$ and $R_{wp}$. The new monoclinic phase gives a better fit to the obtained XRD patterns at above 14.8 GPa. 
\begin{table}[ht!] 
\begin{center}
\begin{tabular}{ |c|c|c|c|c| } 
 \hline
 atom & Wyckoff position & x/a & y/b & z/c \\
 \hline
 Nd & 4e & 0.001(9) & 0.0193(9) & 0.250(6) \\ 
 Fe/Co & 2b & 0.5 & 0.0 & 0.5  \\ 
 Fe/Co & 2c & 0.5 & 0.0 & 0.0  \\
 O & 4e & 0.7986 & 0.4842 & 0.25 \\
 O & 4e & 0.30209 & 0.70552 & 0.95177 \\
 O & 4e & 0.80209 & 0.79448 & 0.04823 \\
 \hline
\end{tabular}\\
\end{center}

 \caption{Relative atomic positions of monoclinic structure at 17.7~\si{\giga \pascal} after Rietveld refinement}
\label{table:2}
\end{table}

\begin{figure}[ht!]
 \flushleft
     \begin{overpic}[width=0.495\linewidth]
     {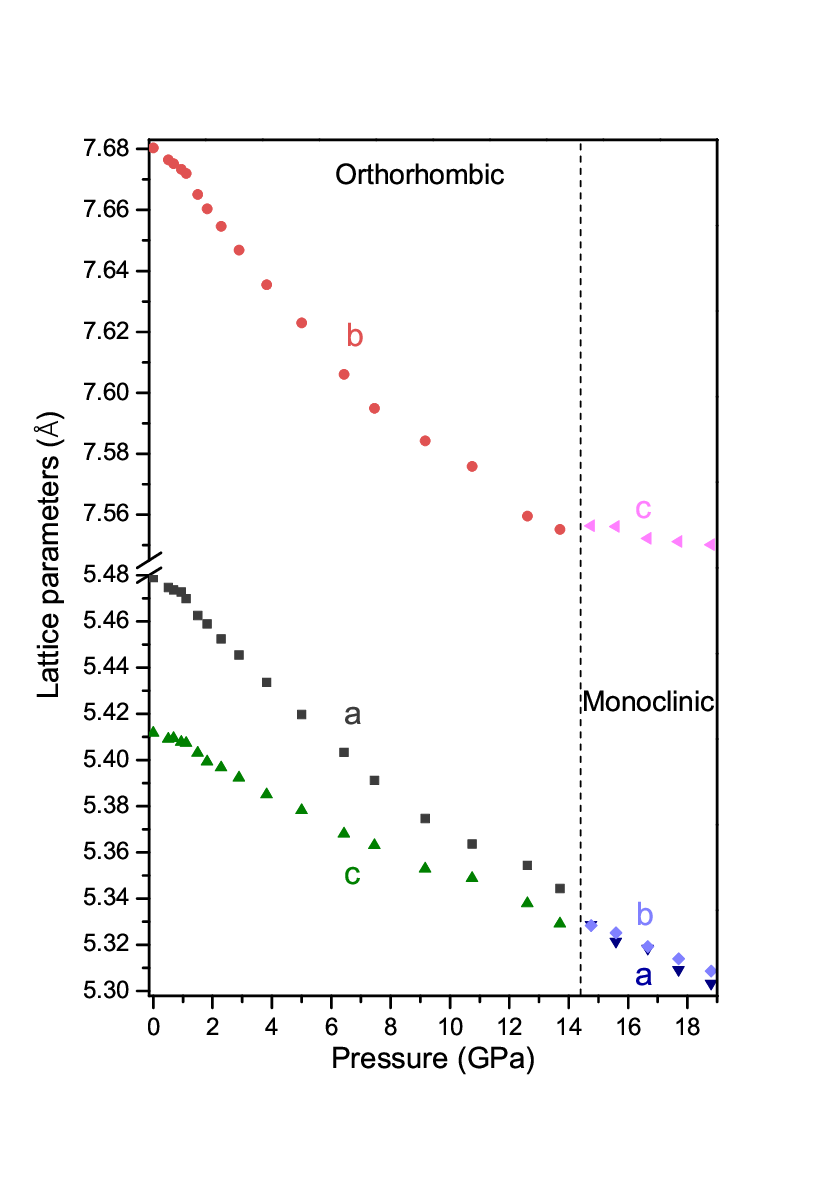} 
     \put (3,95) {(a)}
     \end{overpic}
     \begin{overpic}[width=0.495\linewidth]
     {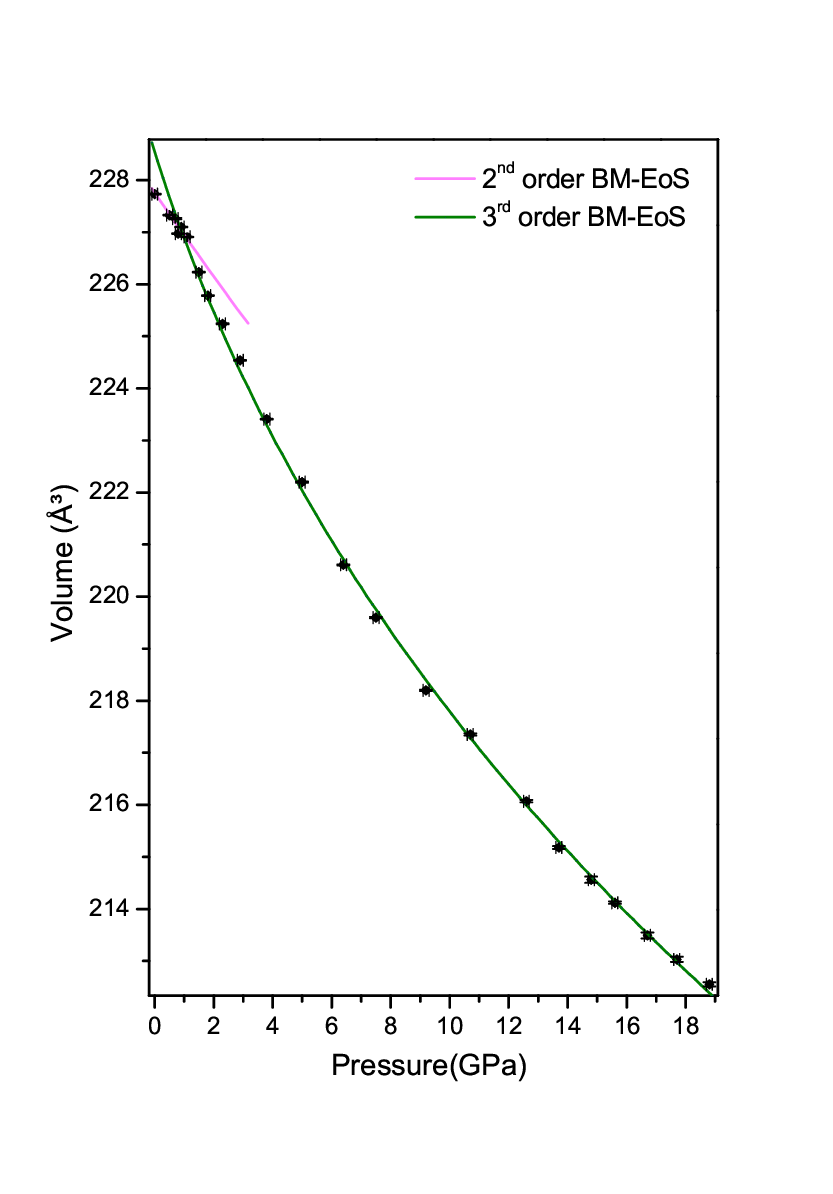}
     \put (3,95) {(b)}
     \end{overpic}
     \caption{Pressure evolution of (a) $a$, $b$ and $c$ lattice parameters and (b) volume of a unit cell of NCFO. The volume data obtained from XRD was represented by solid circles and the solid lines represent BM-EoS fit to the data. }
     \label{fig:Latparam_EoS}
 \end{figure}

To understand structural changes upon applying pressure, the evolution of lattice parameters with pressure is shown in Fig.~\ref{fig:Latparam_EoS}. One can see an anisotropic compression in the lattice parameters with the $a$-axis showing a maximum compression in the orthorhombic phase to about 3\%. Interestingly the volume does not show any observable discontinuity at the orthorhombic to the monoclinic phase transition. A close inspection reveals a distinct slope change in the pressure evolution of lattice parameters and volume at about 1.1~\si{\giga \pascal} in the orthorhombic region even though there is no structural change. We shall first discuss the structural behavior of the orthorhombic phase. For determination of the EoS parameters we have analyzed our volume versus pressure data using Birch Murnaghan Equation of State (BM-EoS) \cite{birch1947,murnaghan1944compressibility}. The volume data for the whole region could not be fitted with a single EoS. Hence, we have carried out two different EoS fits of volume versus pressure data in the two pressure regions, (i) ambient to 1.1~\si{\giga \pascal} and (ii) above 1.1~\si{\giga \pascal}, separately. The volume data up to 1.1~\si{\giga \pascal} fit well to 2$^{\rm nd}$ order BM-EoS, whereas the volume data in between 1.1~\si{\giga \pascal} and 14.8~\si{\giga \pascal} fit well to 3$^{\rm rd}$ order BM-EoS. The Bulk modulus(K$_0$) of these regions are 280(39)~\si{\giga \pascal} and 126.7(5)~\si{\giga \pascal}, respectively. The first-order derivative of Bulk modulus(K$'$) with respect to pressure is found to be 24.0(2) for the pressure region (ii). This large drop in K$_0$ value indicates that the NCFO is less compressible in the lower pressure region (i.e. before 1.1~\si{\giga \pascal}) and above 1.1~\si{\giga \pascal} compressibility increases to a value more than double to that below 1.1~\si{\giga \pascal}. In the absence of an isostructural transition, this can only be explained by the modification in internal bond strengths due to a certain electronic anomaly.

\begin{figure}
     \centering
     \begin{overpic}[width=0.46\linewidth]{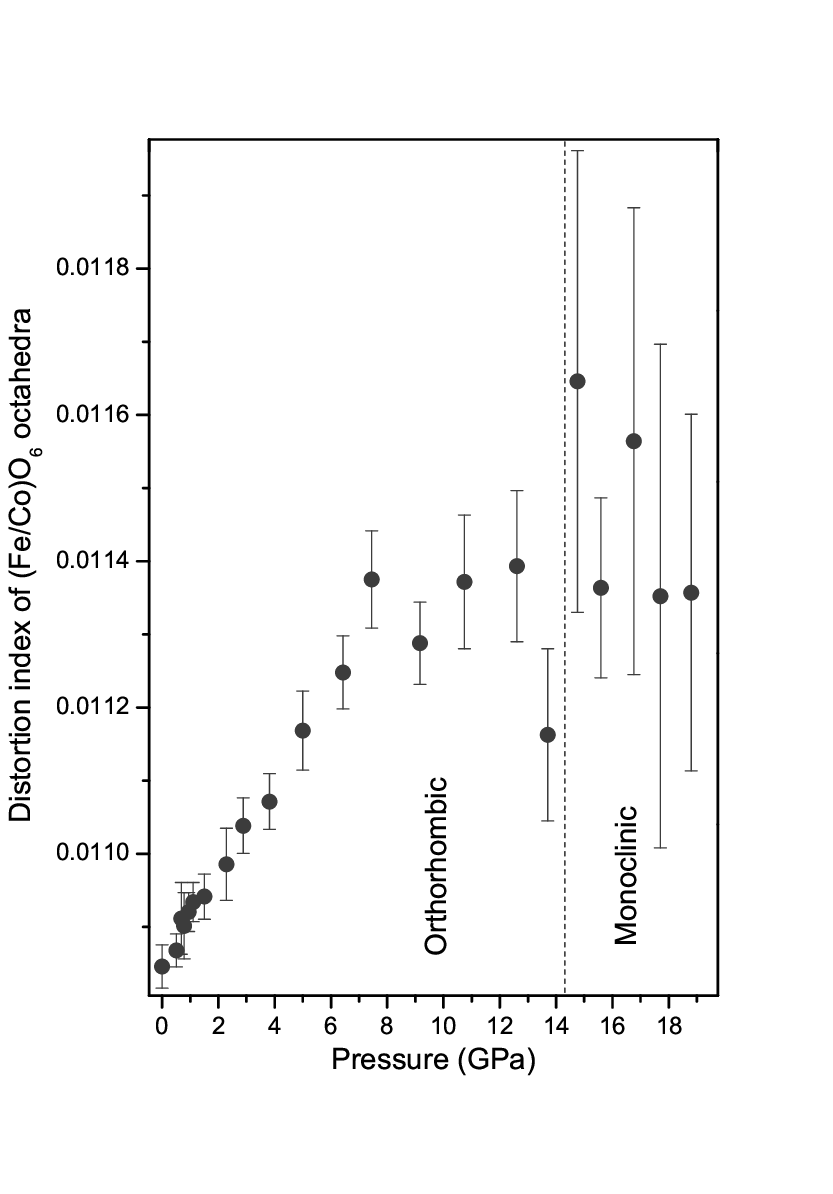}
        \put (3,95) {(a)}
    \end{overpic}
    \begin{overpic}[width=0.46\linewidth]{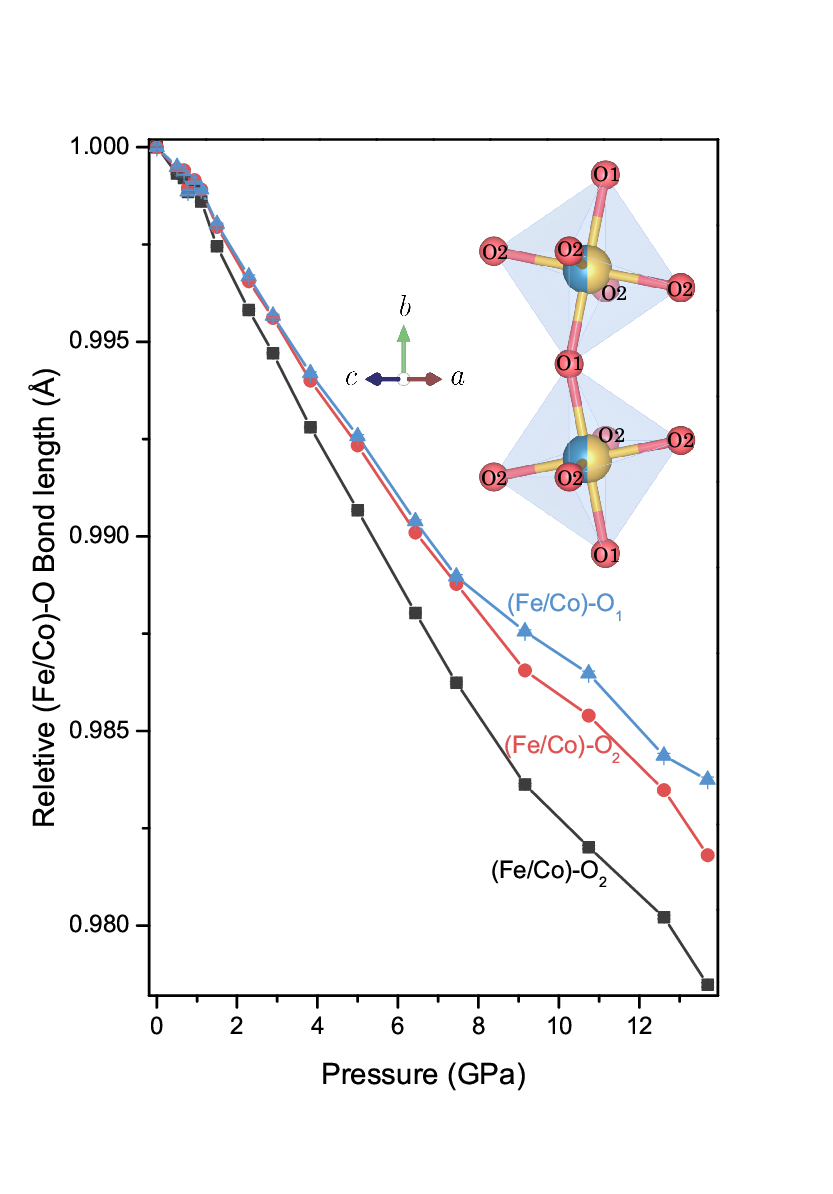}
     \put (3,95) {(b)}
    \end{overpic}
     \caption{(a) Distortion index of (Fe/Co)O$_6$ octahedra with respect to pressure and (b) pressure variation of relative change in different (Fe/Co)-O bond lengths. (Fe/Co)-O1 bond having initial bond-length of 1.97004(3) is along $b$ direction. (Fe/Co)-O2 bond with initial bond-length 1.97168(2) is in the plane of the paper and the one with bond-length 2.01935(3) projects normally through the plane of the paper. They are represented by red circles and black squares respectively.}
     \label{fig:DistInd}
 \end{figure}

From Fig.~\ref{fig:Latparam_EoS}, it is noticeable that after the structural phase transition to the monoclinic phase, the largest cell parameter ($c$-axis) saturate with increasing pressure and the other two cell parameters almost merge with each other. Interestingly, we don't see any observable discontinuity in the volume across the orthorhombic to the monoclinic structural transition. Therefore, we extrapolate the EoS fit of the orthorhombic phase after the structural transition. Extending the EoS fit of the orthorhombic phase, it matches well with the volume up to 18.8~$\si{\giga \pascal}$, the highest pressure of this study. We have calculated the octahedral distortion index(DI) of the (Fe/Co)O$_6$ octahedra.
The distortion index of a (Fe/Co)O$_6$ octahedra can be defined as,

$$DI=\frac{1}{6}\sum_{i=1}^{6}\frac{| l_i − l_{av} |}{l_{av}},$$ where $l_i$ is the individual (Fe/Co)--O bond length, and $l_{av}$ is their average
bond length. The variation of the octahedral $DI$ with pressure is shown in Fig.\ref{fig:DistInd}(a). We find that the distortion index increases with pressure and saturates in the monoclinic phase, with a large oscillatory-type behavior very close to the phase transition pressure. This shows that the internal strain of the octahedra increases in the orthorhombic phase with pressure. The sample undergoes a structural transition to the monoclinic phase to accommodate the large lattice strain and hence to a minimum energy configuration. The structural transition takes place with a change in the space group.

Now we would like to understand the anomaly in the lattice parameters, volume, and compressibility at about 1.1~\si{\giga\pascal} without any structural transition. The bulk modulus drops to less than half of the original value above 1.1~\si{\giga\pascal}. It has been observed that several 3d-transition metal elements and compounds show spin transition at high pressures accompanied by a discontinuity in bulk modulus. High spin to low spin transitions in the 3d-transition metal octahedral complexes take place by depopulation of orbitals leading to a change in ionic radius and hence a change in lattice elastic constants. Electronic spin transitions of Fe$^{3+}$ in the octahedral site of perovskite materials are generally found to occur in a very high-pressure range of about 15 to 80~$\si{\giga\pascal}$ \cite{lin2013,pinku2021}. Vogt et al.~\cite{vogt2003} have shown that Co$^{3+}$ in the octahedral site of perovskite undergoes a pressure-induced intermediate spin (IS) to low spin (LS) transition that completes at about 4~\si{\giga\pascal} where the compressibility decreases considerably. Therefore, we believe that the anomaly in volume compression at about 1~\si{\giga\pascal} is probably due to spin-state transition in Co$^{3+}$ octahedral sublattice. de Oliveira et al. \cite{de2020insulator} have shown that NCFO lattice parameters are strongly dependent on the local magnetic moment of Co. A comparison of lattice parameters shows that in our case NCFO is in a high spin state of Co$^{3+}$. Our structural analysis shows that in $ac$-plane the Nd atoms shift sharply till about 1~\si{\giga\pascal} followed by a sudden change in the slope of the shift with respect to pressure. This leads to a differential compression between Nd-O and Fe/Co-O bonds. (Fe/Co)O$_6$ octahedra have three different Fe/Co-O bond lengths in three directions. In Fig.\ref{fig:DistInd}(b) we have shown pressure evolution of relative change in Fe/Co-O bond lengths, which shows that until about 1~\si{\giga\pascal} they have a similar compression behavior. Above 1~\si{\giga\pascal} the compression increases and it is also anisotropic. NCFO is expected to have strong short-range magnetic interactions at ambient pressure and temperature. Therefore, we believe that above 1~\si{\giga\pascal} the sample undergoes a continuous HS-LS transition due to increased compressibility in Fe/Co-O bond lengths. The differential compressibility of octahedra leads to an increase in octahedral distortion. The Jahn Teller distortion due to the octahedral differential compressibility can affect the local magnetic moment in the sample. But from only XRD investigation it is truly difficult to comment on the magnetic properties of the sample and hence we proceed to further complementary measurements.

\subsection{Raman Spectroscopic Study}
As a complementary study to XRD, we have carried out Raman spectroscopy measurements at ambient, low temperature, high pressure, and with a magnetic field at high pressure. The ambient Raman spectrum matches well with the reported Raman spectrum at room temperature for $\rm Pr_2CoFeO_6$ which is a sister compound of NCFO \cite{pal2019investigation}. According to group theoretical analysis, 24 modes (7A$_g$+5B$_{1g}$+7B$_{2g}$+5B$_{3g}$) are Raman active for $Pnma$ structure\cite{pal2019investigation}.  However, our Raman spectroscopic measurements reveal only two Raman active modes at ambient conditions. Comparing our Raman spectrum with those reported in literature \cite{pal2019investigation,iliev2007modeassignment,LaMnO3modeAssignment}, the strongest mode, 634.4~\si{\centi \meter}$^{-1}$ is assigned to stretching vibrations of (Fe/Co)O$_6$ octahedra (B$_{g}$ symmetry). The second strongest mode at 529.6~\si{\centi \meter}$^{-1}$ is assigned to anti stretching and bending mode of (Fe/Co)O$_6$ octahedra (A$_g$ symmetry). Due to the presence of site disorder of Fe/Co atoms and heavy Nd atoms, the other modes have broadened and are not visible above the background \cite{pal2019investigation}. In fact, the A$_g$ and B$_g$ modes are also quite broad due to the site disorder in this DPO.

\subsubsection{Low-Temperature Raman Analysis}
de Oliveira et al. \cite{de2020insulator} have carried out a detailed study to understand the correlation of electronic, magnetic, and structural behaviour of NCFO. As mentioned in the introduction, the material exhibits an antiferromagnetic transition at about T$_{N}$=246~K and a classical non-interacting paramagnetic state above $T=2.2T_N$ with a strong short-range magnetic interaction in the intermediate temperature range.

\begin{figure}[ht!]
     \includegraphics[width=0.75\linewidth]{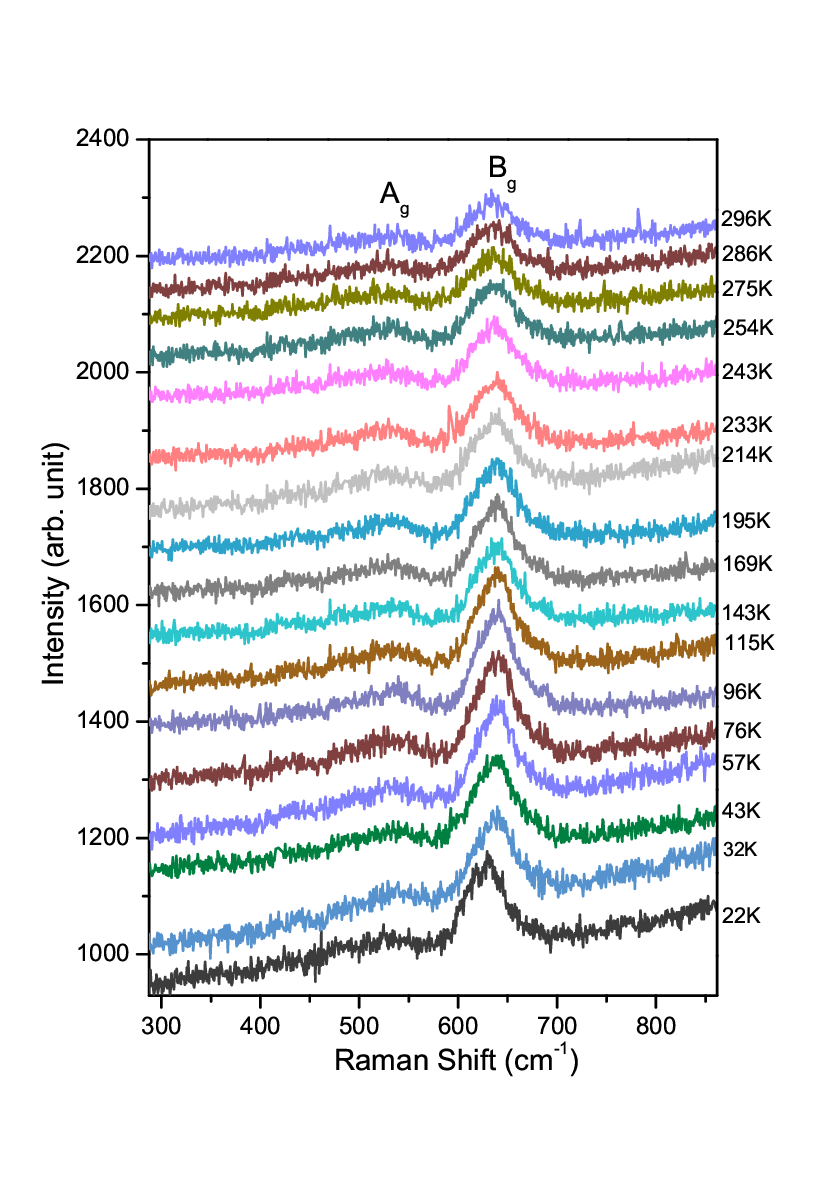}
     \caption{ Evolution of Raman spectra of NCFO with increasing temperature. Temperature is indicated alongside each Raman spectrum. }
     \label{fig:ltRaman}
 \end{figure}

\begin{figure}[t!]
    \centering  
    \begin{overpic}[width=0.48\linewidth]{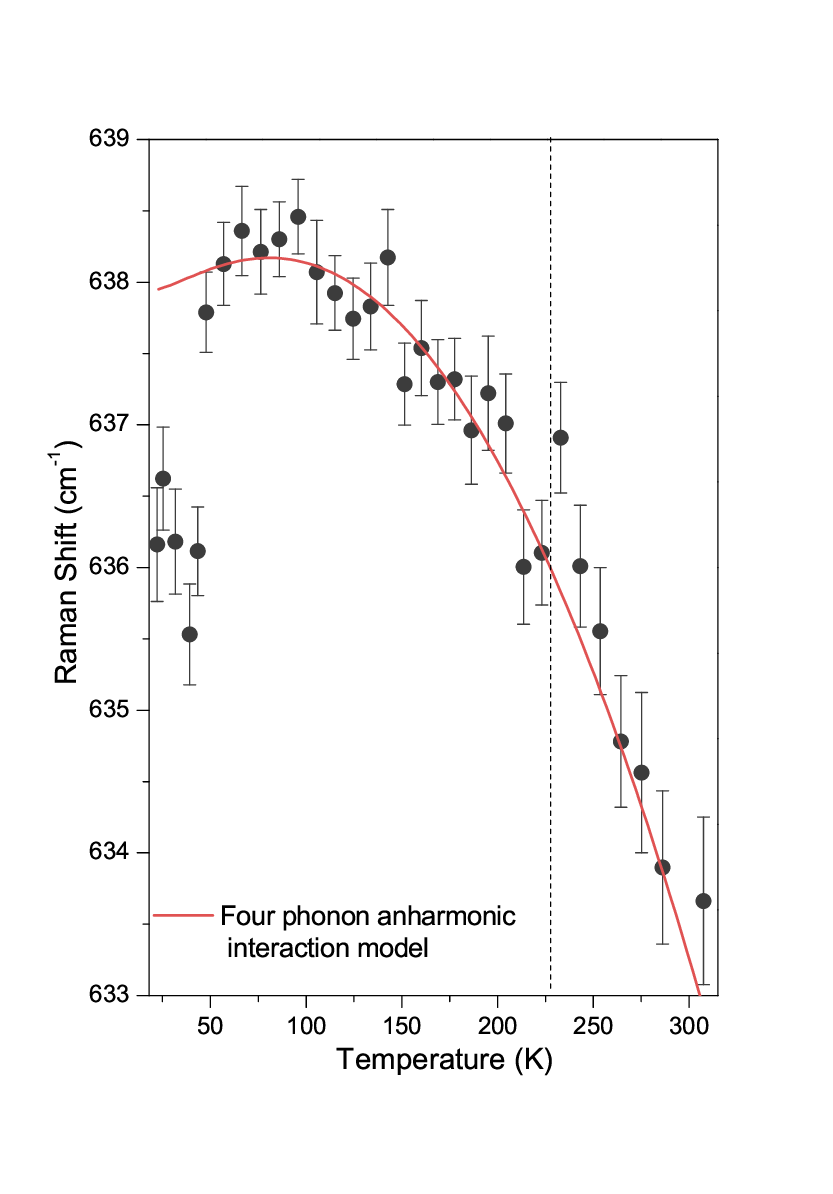}
 \put (1,90) {(a)}
\end{overpic}
\begin{overpic}[width=0.48\linewidth]{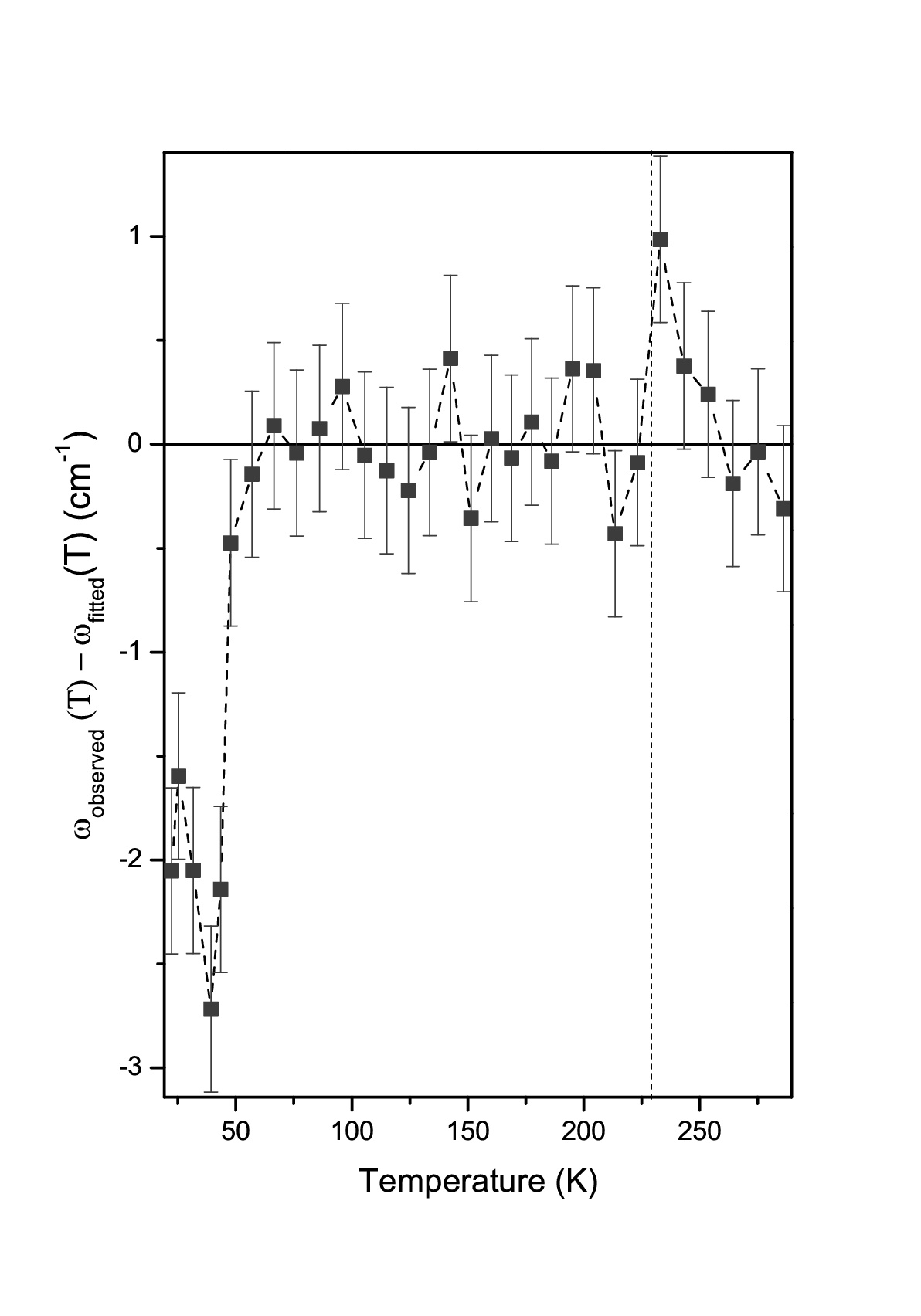}
    \put (1,90) {(b)}
\end{overpic}
    \caption{(a) Variation of Raman shift with temperature. Four phonon interaction model is used to fit the data between 47~K and 223~K and then extrapolated to higher temperatures. (b) Difference between observed $\omega(T)$ and fitted $\omega_{\rm anh}(T)$ Raman shift vs temperature.  }
     \label{fig:ltRamanPeakPos}
 \end{figure}

 Hence we have carried out low-temperature Raman measurements from 22~K to room temperature. In Fig.~\ref{fig:ltRaman}, temperature variations of the Raman spectra are shown at selected temperatures. No new Raman modes were found to appear in the temperature range investigated. To elucidate the variation of the strong B$_{g}$ mode frequency with temperature, the mode is fitted using a Lorentz profile. The mode frequency is plotted against temperature (Fig.~\ref{fig:ltRamanPeakPos}(a)) which shows interesting behaviour. The Raman mode frequency shows a blue shift with a decrease in temperature as expected and then starts saturating below about 100~K. As the temperature decreases further, the Raman mode shows a drastic softening from about 47~K. This softening may be related to the magnetic spin reorientation transition as observed by de Oliveira et al\cite{de2020insulator}.  A similar drop in the (B/B$'$)O$_6$ octahedral stretching Raman mode has been seen in other 3$d$-4$f$ DPO systems due to ferromagnetic transition \cite{iliev2007modeassignment}. A weak ferromagnetic behaviour is observed in the NCFO sample at low temperatures\cite{de2020insulator}. The B$_g$ mode is due to the stretching behaviour of (Fe/Co)O$_6$ octahedra and gives rise to Jahn Teller distortion leading to changes in spin-orbit coupling. Hence the mode is expected to be affected by the magnetic transition, which is clear from the softening.
 
 In the higher temperature region, close observation reveals a sudden drop in the mode frequency below 233~K, which is close to the T$_N$ value. To get a better clarity we have fitted the B$_g$ mode in the temperature range from 223~K to 47~K using four phonon anharmonic interaction model:
  \begin{equation}
 \omega_{\rm anh}(T)=\omega_0-C \left (1+\frac{2}{e^ {\frac{\hbar\omega_0}{2k_BT}}-1} \right )-D \left (1+\frac{3}{e^{\frac{\hbar\omega_0}{3k_BT}}-1}+\frac{3}{\left(e^{\frac{\hbar\omega_0}{3k_{B}T}}-1\right)^2} \right )
 \end{equation}
where, $\omega_0$, C and D are fitting parameters that describe the frequency at 0~K and the strength of three and four phonon processes to the frequency shift, respectively.
The second term in Eq.(\theequation)  corresponds to cubic anharmonicity that describes the decay of one optical phonon into two acoustic phonons of equal frequency. The third term corresponds to the decay of one optical phonon into three identical acoustic phonons\cite{PhysRevBDebabrata}. To obtain any other contribution, we have subtracted  $\omega_{\rm anh}$(T) from the observed Raman shift, $\omega(T)$ and plotted in Fig.~\ref{fig:ltRamanPeakPos}(b). One can see a clear deviation below 47~K and as well as above 240~K. At 240~K, the Raman mode frequency shows a sudden jump and then follows a linear behaviour at higher temperatures. The anomaly at 240~K coincides with the T$_N$ of the sample. Therefore, it can be seen that B$_g$ mode is affected strongly by the change in the magnetic behaviour of the sample. 

\subsubsection{High-Pressure Raman analysis}
\begin{figure}
     \centering
     \begin{overpic}[width=0.75\linewidth]{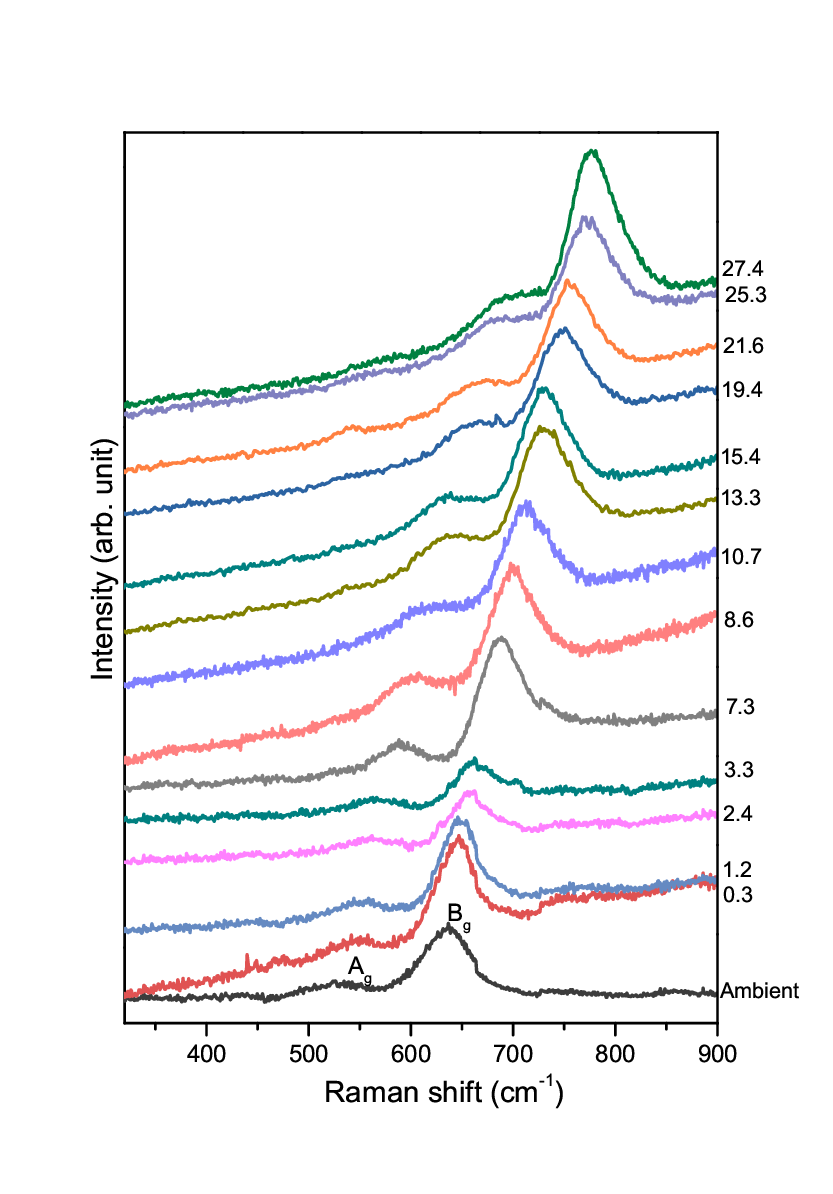}
        
    \end{overpic}
     \caption{Raman spectra of NCFO at some selected pressure points up to 27.4~\si{\giga\pascal}. Pressure (in \si{\giga\pascal}) is indicated alongside each Raman spectrum.}
     \label{fig:hpRaman}
     
 \end{figure}

\begin{figure}
     \centering
     \begin{overpic}[width=0.46\linewidth]{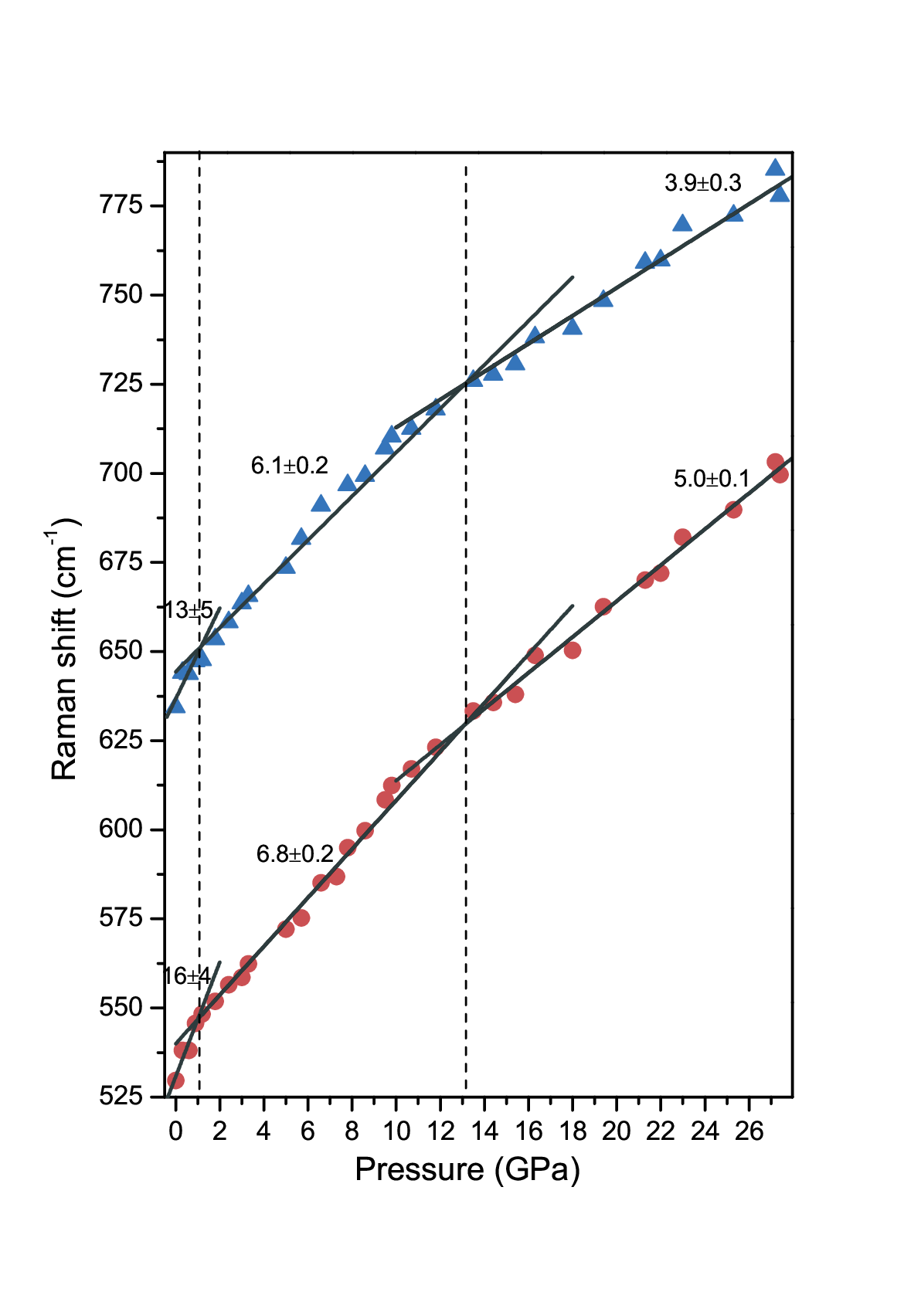}
        \put (3,95) {(a)}
    \end{overpic}
    \begin{overpic}[width=0.46\linewidth]{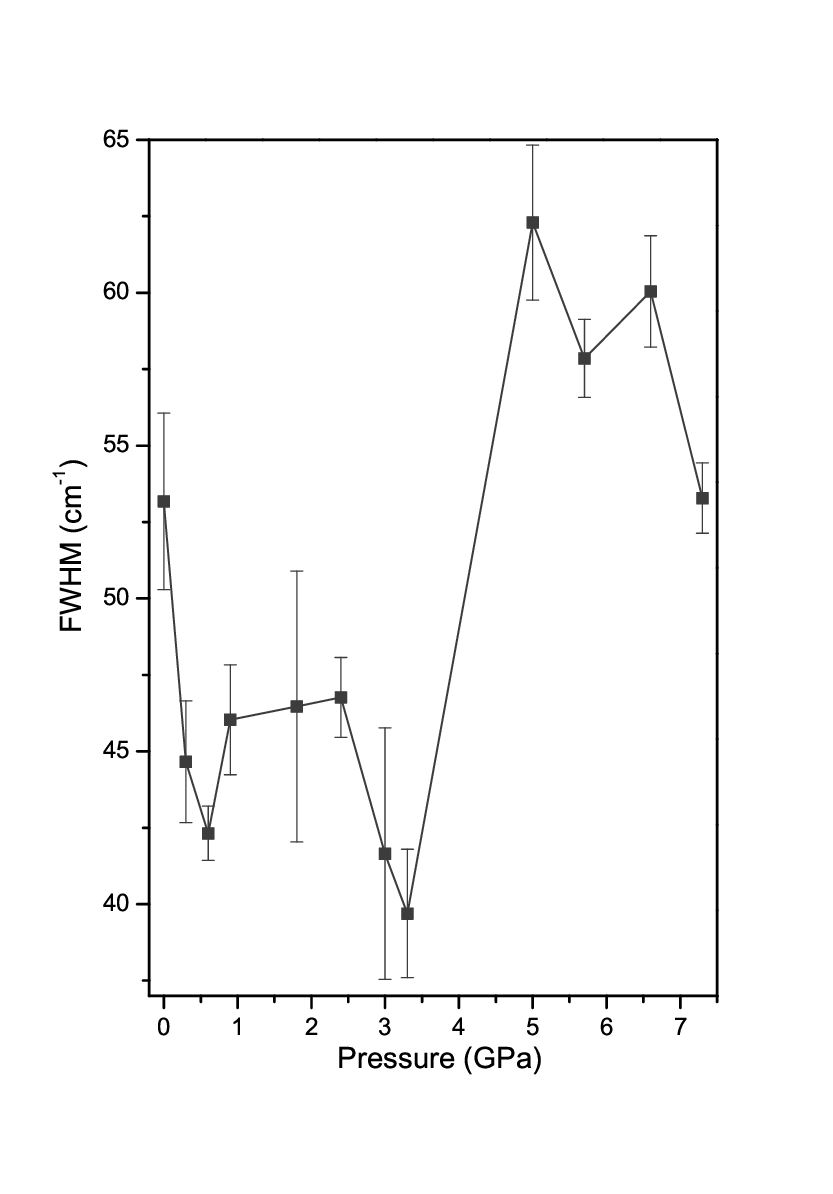}
     \put (3,95) {(b)}
    \end{overpic}
     \caption{(a)  Pressure evolution of Raman peak position reveals two slope changes at around 1~\si{\giga\pascal} and 13.1~\si{\giga\pascal}, (b) FWHM has its minima in a pressure range from about 1~\si{\giga\pascal} to 3.3~\si{\giga\pascal}. }
     \label{fig:peakpos}
 \end{figure}

Pressure evolution of the Raman spectra is investigated in the present study and shown in Fig.~\ref{fig:hpRaman}. We do not see any major change in the Raman spectra up to 27.4~\si{\giga \pascal}, the highest pressure of our study. We have plotted the variation of peak positions of A$_g$ and B$_{g}$ modes with pressure as shown in Fig.~\ref{fig:peakpos}. Both the A$_g$ and B$_{g}$ modes are blue-shifted with increasing pressure and show distinct slope changes at about $\sim1$~\si{\giga \pascal} and $\sim13.1$~\si{\giga \pascal}. Using linear fit, we calculated the slope $(\frac{\mathrm{d} \omega}{\mathrm{d} P})$ corresponding to the A$_g$ and B$_{g}$ modes.
 For the A$_g$ mode, the slope is $16\pm{4}~\si{\centi\metre^{-1}\giga\pascal^{-1}}$ between ambient and $\sim1$~\si{\giga \pascal}. The slope decreases to $6.8\pm{0.2}~\si{\centi\metre^{-1}\giga\pascal^{-1}}$ for the pressure range from $\sim1$~\si{\giga \pascal} to $\sim13.1$~\si{\giga \pascal} and further decreases to $5.0\pm{0.1}~\si{\centi\metre^{-1}\giga\pascal^{-1}}$ beyond $\sim13.1$~\si{\giga \pascal}. For B$_{g}$ mode, the slopes are $13\pm{5}~\si{\centi\metre^{-1}\giga\pascal^{-1}}$, $6.1\pm{0.2}~\si{\centi\metre^{-1}\giga\pascal^{-1}}$ and $3.9\pm{0.3}~\si{\centi\metre^{-1}\giga\pascal^{-1}}$, respectively for the three aforementioned pressure regions. For both modes, the changes in the values of slopes with pressure have a similar trend. The changes in the slopes above 13.1~\si{\giga\pascal} can be related to the orthorhombic to monoclinic structural phase transition. It is interesting to note that the $\mathrm{d} \omega / \mathrm{d} P$ value reduces by more than half above 1.1~\si{\giga\pascal}, even though the compressibility increases. Increasing compressibility indicates an increase in volume change and hence the $\mathrm{d} \omega / \mathrm{d} P$ is expected to increase. The observed reverse effect cannot be related to the elastic softening and may then be due to the electronic effect. The FWHM of B$_g$ mode shows a broad minimum in the pressure region of about 1-3 \si{\giga\pascal}. Since the XRD investigations show an increase in (Fe/Co)O$_6$ octahedral distortion, the Raman mode is expected to broaden further. Instead, the decrease in FWHM shows an increase in phonon lifetime due to a reduction in pressure-induced anharmonic scattering. This can only happen if the pressure drives the sample to a new electronic order. As the sample is insulating the electronic transition can only be related to a new spin ordering in the system.

\begin{figure}
     \centering
     \begin{overpic}[trim=22 43 40 10,clip,width=0.49\linewidth]{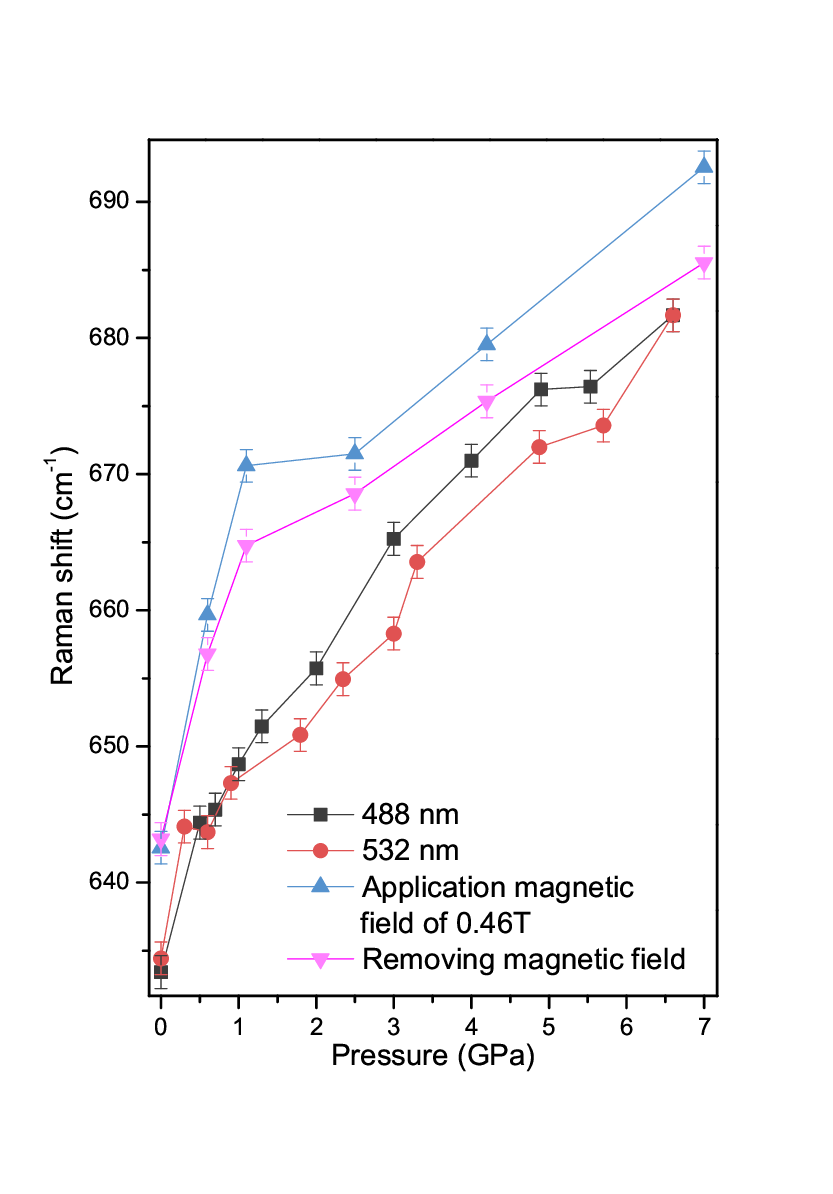}
        \put (3,95) {(a)}
    \end{overpic}
    \begin{overpic}[trim=27 30 40 30,clip,width=0.46\linewidth]{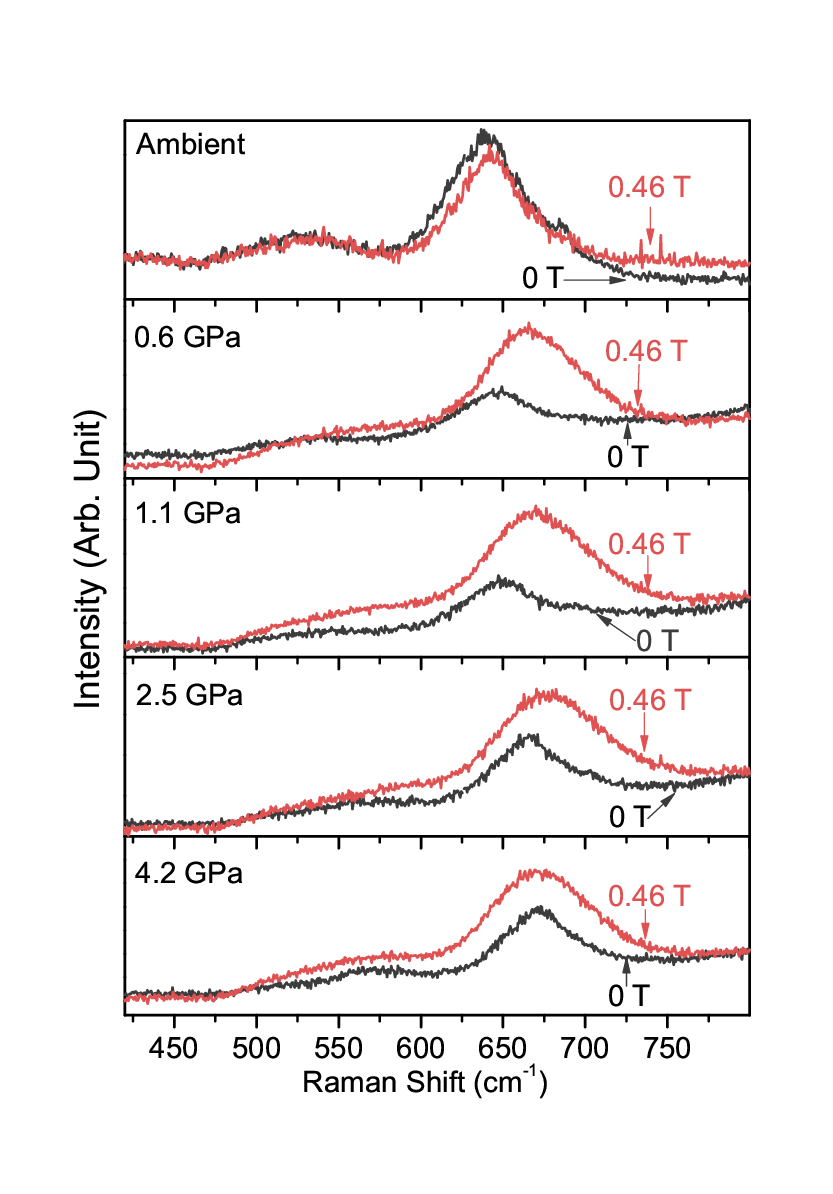}
     \put (3,101) {(b)}
    \end{overpic}
     \caption{(a) Variation of Raman shift with pressure. The black squares and red circles are Raman shift data taken at Horiba (excitation wavelength: 488~\si{\nano\metre}, grating 1800 grooves/mm) and Monovista (excitation wavelength: 532~\si{\nano\metre}, grating 1500 grooves/mm) spectrometers, respectively. The blue (upward pointing) and pink (downward pointing) triangles represent the data taken with the application of a magnetic field of 0.46~T and after withdrawing it, respectively. Both data were taken at Horiba spectrometer with an excitation wavelength of 488~\si{\nano\metre}, (b) Raman spectra at different pressures with and without application of a magnetic field. The pressure is indicated in each plot. The upper (red) and lower (black) plots represent spectra taken with an applied magnetic field of 0.46~T and without any applied magnetic field respectively. }
     \label{fig:Mag}
 \end{figure}

 The anomaly observed at about $\sim1$~\si{\giga \pascal} may correlate with a change in the magnetic behaviour of the sample. Therefore, we have carried out a series of high-pressure Raman experiments using an electromagnet with a maximum magnetic field of about 0.46~T. We have carried out one high-pressure experiment without switching on the magnet up to 6~\si{\giga\pascal} using an excitation laser of wavelength 488~\si{\nano\metre} using the LabRam HR 800 Micro Raman facility. These data were compared with the previous data set taken at 532~\si{\nano\metre} laser using Monovista micro-Raman setup and compared in Fig.~\ref{fig:Mag}. The data looks highly reproducible in this pressure range. The small difference between these two sets is probably due to two different sample loadings, two different spectrometers, two different DACs, and two different excitation wavelengths.
 
 Next, we have carried out our measurements by switching on the magnetic field using Copper-Beryllium DAC. We also collected the Raman spectrum at each pressure point by switching off the magnetic field up to 7~\si{\giga\pascal}. All the data are shown in Fig.~\ref{fig:Mag}. It is interesting to note that at ambient pressure the Raman mode shows a blue shift by 1.5\% with the application of the magnetic field which remained the same even after switching off the magnetic field and waiting for a long time. From Fig.~\ref{fig:Mag}, it can be seen that the Raman mode frequency shows a larger jump up to about 1~\si{\giga\pascal} under the magnetic field followed by a drastic reduction in slope above 1~\si{\giga\pascal}. The Raman mode frequency values on switching off the magnetic field show almost similar jump up to 1.1~\si{\giga\pascal} but then the slope reduces and almost approaches the Raman data taken without any magnetic field. In Fig.~\ref{fig:Mag}(b), we have compared the Raman spectrum taken without a magnetic field and after switching off the magnetic field at certain pressure values. One can see from Fig.~\ref{fig:Mag}(b) that at ambient pressure both the Raman spectra taken with and without magnetic field show excellent match. However, as the pressure is increased the Raman spectra taken under a magnetic field show a considerable broadening in comparison to those taken without a magnetic field. At about 1~\si{\giga\pascal} the B$_g$ mode gets blue shifted by almost 20~\si{\centi\metre}$^{-1}$ when magnetic field is applied. It may also be noted that the Raman spectra do not come back to the original behavior after the magnetic field is switched off. At present, we are unable to explain this interesting phenomenon in NCFO in microscopic detail. In a detailed experimental and theoretical Raman study to understand the magneto-elastic coupling on $\alpha$-SrCr$_2$O$_4$, Duttan et al. \cite{alphaSrCr2O4} have shown that the B$_g$ and A$_g$ modes show a shift of more than 15~\si{\centi\metre}$^{-1}$ due to spin-phonon coupling. They have also shown that the FWHMs of these modes show extensive broadening. Therefore it may be possible that in NCFO application of a magnetic field enhances the spin-phonon interaction due to the presence of short-range ferromagnetic interaction and results in the blue shift as well as broadening of the Raman mode. This experiment shows that there is a strong spin-phonon interaction that affects this B$_{g}$ mode. The hardening of Raman mode frequency, caused by the application of a magnetic field, is possibly due to the strong interaction between the spins and the magnetic field which affects the spin-phonon interaction. By increasing pressure, reduction in volume leads to enhanced magnetic interactions which possibly leads to further hardening of the Raman mode. Above 1~\si{\giga\pascal} the sample probably settles down gradually to a classical paramagnetic state which leads to the softening in bulk modulus and hence reduction in the slope of the Raman mode with respect to pressure. Therefore, we believe that the anomaly at 1~\si{\giga\pascal} is due to the magnetic instability in the system. Our experiments reveal that disordered DPOs are very much affected by lattice strain due to strong magneto-elastic coupling. Further detailed theoretical studies are needed to understand these effects in microscopic detail.

	\section{Conclusions}
	In this study, we have carried out a detailed investigation of the high-pressure behavior of a disordered double perovskite oxide, NCFO, using XRD and Raman measurements. Under high pressure, NCFO undergoes a structural phase transition from the orthorhombic phase to the monoclinic phase at around 13.8~\si{\giga\pascal}, accompanied by significant changes in lattice parameters. The distortion index of the (Fe/Co)O$_6$ octahedra increases with pressure in the orthorhombic phase and shows oscillatory nature around the pressure of phase transition. This indicates that the internal strain of the octahedra increases with pressure, leading to a structural transition to the monoclinic phase to accommodate the large strain.
  Our Raman spectroscopy reveals two active Raman modes---A$_g$ and B$_{g}$. A comparative analysis of the low-temperature Raman spectrum, with the literature, unveils that magnetic interactions and spin reorientation transitions strongly influence the behavior of the B$_{g}$ Raman mode. In high-pressure Raman measurements, both modes exhibit blue shifts with pressure, displaying slope changes at around 1~\si{\giga\pascal} and 13.1~\si{\giga\pascal}. A decrease in the slope of Raman modes with increasing pressure, together with an anomalous decrease in bulk modulus at around 1.1~\si{\giga\pascal}, without any significant structural transition, is observed, possibly related to a new spin ordering in the system. The B$_g$ mode hardens with pressure when subjected to an external magnetic field. The observed anomalies and phase transitions under pressure indicate a rich interplay between lattice and magnetic behavior, making  Re$_2$BB$'$O$_6$ family a promising candidate for further investigations in magneto-structural coupling.

	\section{Acknowledgments}
 The authors acknowledge the financial support from the Department of Science and Technology, Government of India, for the collaboration with XPRESS beamline in the ELETTRA
Synchrotron light source under the Indo-Italian Executive Programme of Scientific and Technological Cooperation. BM acknowledges the UGC, Government of India, for the
financial support to carry out the PhD work.

 \bibliography{manuscript}
\newpage
\section{Figures}
\vspace{-1.5cm}


\end{document}